\documentclass[aps,prl,twocolumn,groupedaddress,amssymb,amsfonts,showpacs]{revtex4}
\usepackage{times,color}

\newcommand{\ignore}[1]{}
\newcommand{\mComment}[1]{}
\newcommand{\lComment}[1]{}
\renewcommand{\mComment}[1]{\textcolor{blue}{Manny: #1}}
\renewcommand{\lComment}[1]{\textcolor{red}{Lorenza: #1}}

\def\cc{{\cal C}}
\def\cd{{\cal D}}

\def\cg{{\cal G}}
\def\ch{{\cal H}}
\def\cs{{\cal S}}
\def\ct{{\cal T}}
\def\cp{{\cal P}}
\def\cq{{\cal Q}}

\def\cj{{\cal J}}

\def\ct{{\cal T}}
\def\cu{{\cal U}}

\def\cz{{\cal Z}}

\newcommand{\cmplxs}{{\mathbb C}}
\newcommand{\nintg}{{\mathbb N}}

\begin{document}


\title{  Robust dynamical decoupling with bounded controls }



\author{ Lorenza Viola }
\email{ lviola@lanl.gov } 
\author{Emanuel Knill}
\email{knill@lanl.gov}
\affiliation{ Los Alamos National Laboratory, Mail Stop B256, 
Los Alamos, New Mexico 87545 }



\date{July 15, 2002}

\begin{abstract}
We propose a general procedure for implementing dynamical decoupling
without requiring arbitrarily strong, impulsive control actions. 
This is accomplished by designing continuous decoupling propagators
according to Eulerian paths in the decoupling group for the 
system.  Such Eulerian decoupling schemes offer two important advantages 
over their impulsive counterparts: they are able to enforce the 
same dynamical symmetrization but with more realistic control resources and, 
at the same time, they are intrinsically tolerant against a large class 
of systematic implementation errors.
\end{abstract}



\maketitle


Dynamical decoupling provides a well-defined framework for
addressing a variety of issues associated with the manipulation of
open quantum systems and interacting quantum subsystems. 
Inspired by coherent averaging methods in nuclear magnetic
resonance spectroscopy~\cite{haeberlen:qc1968}, and cast in 
control-theoretic terms in~\cite{viola:qc1998a,viola:qc1999a},
decoupling techniques are attracting growing interest from the 
quantum control and quantum information processing (QIP) communities.  
Significant applications have resulted in the area of reliable QIP, 
where decoupling has been instrumental in the development 
of quantum error suppression 
schemes~\cite{viola:qc1999a,zanardi:qc1999a,viola:qc1999b},
with the potential for noise-tolerant universal quantum
computation on dynamically generated noiseless
subsystems~\cite{viola:qc2000b}.  In addition, variants of the basic
decoupling concepts play a role in protocols for universal quantum
simulation of both closed- and open-system
dynamics~\cite{wocjan:qc2001a,dodd:qc2002a,lloyd:qc2002a}, with 
implications for encoded simulation~\cite{wu:qc2001a,viola:qc2002a}.  
In a broader context, applications of dynamical decoupling 
to problems that range from inhibiting the decay of unstable 
states~\cite{agarwal:qc2001a}, to suppressing magnetic state 
decoherence~\cite{berman:qc2000a}, or reducing heating effects in 
linear ion traps~\cite{vitali:qc2001a} have been recently envisaged.

From the point of view of implementation, dynamical decoupling 
has relied on the ability of effecting sequences of arbitrarily
strong, instantaneous control pulses.  That is, it required the
ability to impulsively apply a set of control Hamiltonians 
with unbounded strength (the {\it bang-bang} (b.b.)
assumption~\cite{viola:qc1998a}).
While providing a convenient starting point, such a scenario
suffers from being extremely unrealistic for applications.  
In a physical control setting, additional disadvantages associated 
with b.b. decoupling include the difficulty of simultaneously
describing the evolution under the natural (drift) Hamiltonian and 
the control terms, as well as the poor spectral selectivity of
b.b. pulses, with substantial off-resonance effects. Finally,
although compensation techniques based on composite rotations 
exist for stabilizing control pulses against operational
imperfections~\cite{tycko:1983}, they are hard to reconcile with 
the b.b. framework, which does not easily lend itself to incorporating
robustness features. 

In this Letter, we overcome the shortcomings of the b.b. formulation 
by showing how to implement dynamical decoupling based on continuous
modulation of bounded-strength Hamiltonians. If ${\cg}$ is the
discrete group specifying the desired b.b. decoupler, the basic idea 
is to constrain the motion of the control propagator during each cycle 
along a path that interpolates between the elements of ${\cg}$. 
Under mild assumptions on the control Hamiltonians, a decoupling 
prescription inducing the same symmetry structure as in the b.b. 
limit can be constructed by exploiting {\it Eulerian cycles on a 
Cayley graph of} ${\cg}$.  In addition to significantly weakening 
the relevant implementation requirements, Eulerian decoupling turns 
out to be largely insensitive to control faults, opening the way to 
the robust dynamical generation of noise-protected subsystems.

{\it Decoupling setting.$-$} Let the target system $S$ be defined on a
finite-dimensional state space $\ch_S$, and let End$(\ch_S)$ be
the corresponding operator algebra. Thus, $\ch_S \simeq \cmplxs^d$,
End$(\ch_S)\simeq \text{Mat}_d(\cmplxs)$ for some $d$, with $d=2^n$ 
for an $n$-qubit system.  $S$ may be coupled to an uncontrollable 
environment $E$, whereby the evolution on the joint state space
$\ch_S \otimes \ch_E$ is ruled by a total drift Hamiltonian 
$H_0 = H_S \otimes \openone_E + \openone_S \otimes H_E +
\sum_\alpha S_\alpha \otimes E_\alpha$ for appropriate traceless
noise generators $S_\alpha \in \text{End}(\ch_S)$~\cite{viola:qc1999a}. 
A decoupling problem is concerned with characterizing the 
effective evolutions that can be generated from $H_0$ via the 
application of a control field $H_c(t)\otimes \openone_E$ acting 
on $S$ alone~\cite{viola:qc1999a}. Let the control propagator be 
\begin{equation}
U_c(t) = \ct\hspace{-.6mm}\exp\Big\{ \hspace{-.6mm}
-i \int_0^t dt' \, H_c(t') \Big\} \:,
\label{Uc}
\end{equation} 
with $\hbar=1$. In a frame that removes the control field, the 
dynamics is governed by a time-dependent Hamiltonian 
$\tilde{H}(t) $$= U_c^\dagger (t) H_0 U_c(t)$, and the overall  
evolution in the Schr\"{o}dinger picture results from the net 
propagator
\begin{equation}
U (t) = U_c(t) \ct\hspace{-.5mm}\exp\Big\{ \hspace{-.5mm}
-i \int_0^t dt' \, \tilde{H}(t') \Big\} \:.
\label{Us}
\end{equation} 
Assuming that the control action is cyclic, $U_c(t+T_c)=U_c(t)$ for 
some cycle time $T_c >0$ and for all $t$, the stroboscopic dynamics 
$U(t_M=MT_c)$, $M \in \nintg$, can be identified with
the effective evolution induced by $\tilde{H}(t)$ in (\ref{Us}).
First-order decoupling aims at 
generating the desired evolution to lowest order in 
$T_c$, $U(t_M) = \exp(-i \overline{H}^{(0)} t_M)$, where 
\begin{equation}
\overline{H}^{(0)} = {1 \over T_c} \int_0^{T_c} 
dt' \, U_c^\dagger(t') {H_0} U_c(t') \:.
\label{avH}
\end{equation} 
While higher-order corrections can be systematically evaluated, the 
approximation (\ref{avH}) tends to become exact as the fast control 
limit $T_c \rightarrow 0$ is 
approached~\cite{haeberlen:qc1968,viola:qc1999a,viola:qc1999b}. 

In the simplest b.b. decoupling setting, the time-average in 
(\ref{avH}) maps directly into a group-theoretical average. 
Let $\cg$ be a discrete group of order $|\cg|>1$, $\cg=\{ g_j\}$, 
$j=0,\ldots,$ $|\cg| -1$, acting on $\ch_S$ via a faithful, unitary,
projective representation $\mu$, $\mu(\cg) \subset \cu(\ch_S)$.  
Let images of abstract quantities under $\mu$ be denoted as 
$\mu(g_j)=\hat{g}_j$, and so forth~\cite{note}. 
Then b.b. decoupling according to $\cg$ is 
implemented by specifying $U_c(t)$ over each of the $|\cg|$ equal 
sub-intervals defining a control cycle~\cite{viola:qc1999a}:
\begin{equation}
U_c\Big((\ell-1) \Delta t + s\Big) = \hat{g}_{\ell-1}\:, \hspace{5mm}
s \in [0, \Delta t) \:,
\label{bb}
\end{equation}
with $T_c=|\cg| \Delta t$ for $\Delta t >0$, and $\ell=1,\ldots, 
|\cg|$. The resulting control action corresponds to extracting the 
$\cg$-invariant component of $H_0$, 
$\overline{H}^{(0)}=\Pi_{\widehat{\cg}}(H_0)$, where 
\begin{equation}
\Pi_{\widehat{\cg}}(X)= {1 \over |\cg|}
\sum_{g_j \in \cg} \hat{g}_j^\dagger\, X \,\hat{g}_j \:,
\hspace{5mm} X \in \text{End}(\ch_S)\:, 
\label{piG}
\end{equation}   
is the projector onto the commutant $\widehat{\cmplxs \cg}$$'$ of 
$\widehat{\cmplxs \cg}$ in 
End($\ch_S)$~\cite{viola:qc1999a,zanardi:qc1999a}. 
According to (\ref{bb}), $U_c(t)$ 
jumps from $\hat{g}_{\ell -1}$ to $\hat{g}_{\ell}= 
(\hat{g}_{\ell}\hat{g}_{\ell -1}^\dagger) \hat{g}_{\ell -1}$ through 
the application of an arbitrarily strong, instantaneous control kick 
at the $\ell$'th endpoint $t_\ell= \ell \Delta t$, realizing the b.b. 
pulse $p_\ell= \hat{g}_{\ell}
\hat{g}_{\ell -1}^\dagger$~\cite{viola:qc1999b}. 

{\it Eulerian dynamical decoupling.$-$} We seek a way for smoothly 
steering $U_c(t)$ from $\hat{g}_{\ell -1}$ to $\hat{g}_{\ell}$ by a 
control action distributed along the whole $\ell$'th sub-interval. 
Let ${\Gamma} = \{ {\gamma}_\lambda \}$, $\lambda=1, \ldots, |\Gamma|$ 
be a generating set for ${\cg}$. The {\it Cayley graph} 
$G({\cg}, {\Gamma})$ of ${\cg}$ with respect to ${\Gamma}$ is the
directed multigraph whose edges are coloured with the 
generators~\cite{bollobas:1998}, where vertex ${g}_{\ell -1}$ is 
joined to vertex ${g}_\ell$ by an edge of colour $\lambda$ if and 
only if ${g}_{\ell}{g}_{\ell -1}^{-1} = {\gamma}_\lambda$ 
{\it i.e.}, ${g}_{\ell}=  {\gamma}_\lambda {g}_{\ell -1}$.  
Physically, imagine that we have the ability to implement each 
generator $\hat{\gamma}_\lambda$, by the application of 
control Hamiltonians $h_\lambda(t)$ over $\Delta t$, 
\begin{equation}
\hat{\gamma}_\lambda = \ct\hspace{-.6mm}\exp\Big\{ 
\hspace{-.6mm}-i \int_0^{\Delta t} \hspace{-.5mm}dt' \, 
h_\lambda (t') \Big\} \:, 
\hspace{4mm} \lambda=1, \ldots, |\Gamma|\:.
\label{gen}
\end{equation}
The choice of $h_\lambda(t)$ is not unique, allowing for
additional implementation flexibility. Once a choice is made, the
control action is determined by assigning a cycle time and a rule 
for switching the Hamiltonians $h_\lambda(t)$ during the cycle
sub-intervals. We show how a useful rule results from sequentially
implementing generators so that they follow a {\it Eulerian cycle} 
on $G({\cg}, {\Gamma})$. A Eulerian cycle is defined as a cycle
that uses each edge exactly once~\cite{bollobas:1998}.  Because a
Cayley graph is regular, it always possesses Eulerian cycles, whose 
length is necessarily $L= |\cg||\Gamma|$~\cite{bollobas:1998}.

Let a Eulerian cycle beginning at the identity $g_0$ of $G$
be given by the sequence of edge colours used, $\cp_E=(p_\ell)_\ell$, 
with $\ell=1,\ldots, L$, and $p_\ell= {\gamma}_\lambda$ for some 
$\lambda$, for every $\ell$. Note that each vertex has exactly
one departing edge of each colour, so that $\cp_E$ determines
a well defined path. 
We define Eulerian decoupling according to $\cg$ by letting 
$T_c=L \Delta t$ and by assigning $U_c(t)$ as follows:
\begin{equation}
U_c\Big((\ell-1) \Delta t + s\Big) = u_\ell (s) \,
U_c\Big((\ell-1) \Delta t \Big) \:, 
\label{euler}
\end{equation}
where $s \in [0,\Delta t)$, and 
$u_\ell (s)=\ct \hspace{-.4mm}\exp(\hspace{-.6mm}-i \int_0^{s} 
\hspace{-.5mm}dt' \, h_\ell (t') )$, $u_\ell (\Delta t)=\hat{p}_\ell$, 
$\ell=1,\ldots,L$. This decoupling prescription means that during 
the $\ell$'th sub-interval one chooses as a control Hamiltonian the 
one that implements the generator $\hat{\gamma}_\lambda$, with
$\gamma_\lambda$ colouring the edge 
$p_\ell$ in $\cp_E$. The effective Hamiltonian $\overline{H}^{(0)}$ 
under Eulerian decoupling is obtained by evaluating the time average 
(\ref{avH}) based on (\ref{euler}). The resulting $L$ terms can be 
partitioned into $|\Gamma|$ families, each corresponding to a fixed 
generator ${\gamma}_\lambda$. Because $\cp_E$ contains exactly 
one ${\gamma}_\lambda$-coloured edge ending at any given 
vertex ${g}_j$, each family effects a sum over the group elements
as in (\ref{piG}). Thus, the quantum operation $\cq_{\widehat{\cg}}$ 
corresponding to (\ref{euler}) can be expressed as 
\begin{equation}
\cq_{\widehat{\cg}}(X) = \Pi_{\widehat{\cg}} 
( F_{\widehat{\Gamma}} (X)) \:, 
\hspace{3mm} X \in \text{End}(\ch_S) \:, 
\label{qG}
\end{equation}
with the map $F_{\widehat{\Gamma}}$ implementing an average over both 
the group generators and control sub-interval:
\begin{equation}
F_{\widehat{\Gamma}}(X) = {1 \over |\Gamma|} \sum_{\lambda=1}^{|\Gamma|}
{1 \over \Delta t} \int_0^{\Delta t} \hspace{-.8mm} ds \:
u_\lambda^\dagger (s) X u_\lambda (s) \:.
\label{F} 
\end{equation}

Thanks to the way $\Pi_{\widehat{\cg}}$ enters (\ref{qG}), 
$\cq_{\widehat{\cg}}(X) $$\in \widehat{\cmplxs \cg}$$'$ for an 
arbitrary input $X$. This property will be repeatedly used in the 
following. The link between Eulerian decoupling and $\cg$-symmetrization 
is established upon enforcing some additional compatibility between 
$\Pi_{\widehat{\cg}}$ and $F_{\widehat{\Gamma}}$. 

{\it Theorem.} Let $X$ be any (time-independent) operator on $\ch_S$,
and let $\cq_{\widehat{\cg}}$ be defined as above. If the controls 
are chosen in the decoupling group algebra, $h_\ell(t) \in 
\widehat{\cmplxs \cg}$ for all $t \in [0,\Delta t]$ and for all 
$\ell=1,$$\ldots L$, then 
$$ \cq_{\widehat{\cg}}(X) = \Pi_{\widehat{\cg}} (X)\:, 
\hspace{3mm} X \in \text{End}(\ch_S) \:. $$

{\it Proof.} The assumption on the controls
implies that $u_\lambda (s) \in \widehat{\cmplxs \cg}$ 
$\forall \lambda$, $\forall s\in [0,\Delta t]$. Thus, 
$ F_{\widehat{\Gamma}}(Y)=Y$ for every (time-independent) 
$Y \in  \widehat{\cmplxs \cg}$$'$.
Now let $X \in \text{End}(\ch_S)$ and calculate
$\cq^2_{\widehat{\cg}}(X) = $
$\Pi_{\widehat{\cg}}(\cq_{\widehat{\cg}} (X))$ $= \Pi_{\widehat{\cg}}
(F_{\widehat{\Gamma}}(X))= \cq_{\widehat{\cg}}(X)$. Thus, 
$\cq_{\widehat{\cg}}$ is a projector. 
Because $\text{Range}\,\cq_{\widehat{\cg}} \subseteq  
\widehat{\cmplxs \cg}$$'$, 
$\cq_{\widehat{\cg}} = \Pi_{\widehat{\cg}}$  iff $\cq_{\widehat{\cg}}$ 
has identity action on $\widehat{\cmplxs \cg}$$'$. 
Let $Y \in \widehat{\cmplxs \cg}$$'$, then 
$\cq_{\widehat{\cg}} (Y) =\Pi_{\widehat{\cg}}(F_{\widehat{\Gamma}}(Y)) $ 
$= \Pi_{\widehat{\cg}}(Y)$. 
\hspace*{\fill}\mbox{\rule[0pt]{1.4ex}{1.4ex}}

The b.b. limit is formally recovered by letting
$F_{\widehat{\Gamma}}$ be the identity map. In the Eulerian approach,
at the expense of lengthening the control cycle by a factor of
$|\Gamma|$, the same $\cg$-symmetrization can be attained using
{\it bounded} controls. The maximum strengths achievable in implementing 
the generators (\ref{gen}) directly affects the minimum attainable $T_c$,
and therefore the accuracy of the averaging~\cite{viola:qc1999a}. 
While the overhead $|\Gamma|$ depends
on the specific group, it is worth noting that, similar to
$\Pi_{\widehat{\cg}}$~\cite{zanardi:qc1999a}, $\cq_{\widehat{\cg}}$
satisfies the property that $\cq_{\widehat{\cg}}(X)=
\cq_{\widehat{\cg/\cg_0}}(X)$ whenever $\cg_0$ is a normal subgroup of
$\cg$ and $X \in \widehat{\cmplxs \cg_0}$$'$~\cite{note2}. 
Thus, if the dynamics is already
$\cg_0$-invariant, Eulerian decoupling according to $\cg$ can be
accomplished by using a Cayley graph of the smaller quotient group
$\cg/\cg_0$.

{\it Robustness analysis.-} The fact that control actions are 
now distributed along finite time intervals 
translates into major gains in terms of resilience of Eulerian 
schemes against imperfections in the controls themselves. Imagine 
that systematic implementation errors result in a faulty 
control Hamiltonian $H'_c(t)$, and partition $H'_c(t)$ into
\begin{equation}
H'_c(t)= H_c(t) + \Delta H_c(t) \:,
\label{part}
\end{equation}
such that $H_c(t)$ $\in $$\widehat{\cmplxs \cg}$ is the intended control   
Hamiltonian, and $\Delta H_c(t)$ is the error component. Now work in 
the same frame used earlier, which only removes the ideal control part 
from the effective Hamiltonian. 
Because $H(t)= H_0+H'_c(t)=[H_0+\Delta H_c(t)]+H_c(t)$, this maps the 
evolution under $H_0$ with the faulty control $H_c'(t)$ into the 
evolution under $H_0+ \Delta H_c(t)$ with the ideal control. Thus, 
the new effective dynamics is obtained by replacing $H_0$ with 
$H_0+\Delta H_c(t)$ in (\ref{avH}). 

Suppose that the faults are properly correlated with the underlying 
path, meaning that every time a particular generator 
$\hat{\gamma}_\lambda$ is implemented, the same imperfection occurs
at equivalent temporal locations within the sub-interval, regardless 
of the position of ${\gamma}_\lambda$ along $\cp_E$. Then 
$\Delta H_c ((\ell-1) \Delta t + s) = \Delta h_\lambda (s)$, $\lambda$ 
being the colour of the edge that $\cp_E$ uses during the $\ell$'th 
sub-interval. By a similar calculation as in the ideal case, the 
quantum operation $\cq_{\widehat{\cg}}$ is modified as follows:
\begin{equation}
\cq_{\widehat{\cg}}'(X) = \Pi_{\widehat{\cg}}(X) + 
\cq_{\widehat{\cg}}(\Delta H_c)\:, \hspace{3mm} X 
\in \text{End}(\ch_S) \:,
\end{equation}
where $\cq_{\widehat{\cg}}(\Delta H_c)$ can be computed as in (\ref{qG}) 
and (\ref{F}), but with the operator $X$ in the integral replaced by one 
that depends on $s$ and $\lambda$. Thus, $\cq_{\widehat{\cg}}(\Delta H_c)$ 
is a functional of the fault history over $[0,\Delta t]$, which 
characterizes the residual control errors experienced by the system.
Notably, two useful features emerge: without extra assumptions, such 
residual control errors belong to $\widehat{\cmplxs \cg}$$'$. If, 
in addition, $\Delta H_c(t)$ is itself (as $H_c(t)$) in 
$\widehat{\cmplxs \cg}$, then all control effects remain in 
$\widehat{\cmplxs \cg}$, and the residual control errors belong to 
the center $Z_{\widehat{\cmplxs \cg}} =
\widehat{\cmplxs \cg} \cap \widehat{\cmplxs \cg}$$'$. 
 
The effects of $\cq_{\widehat{\cg}}(\Delta H_c)$ may still adversely 
impact the performance of the system. However, they can be compensated 
for by encodings in appropriate subsystems~\cite{viola:qc2000b}. 
Let $J\in \cj$ label the irreducible components of 
$\widehat{\cmplxs \cg}$. Then $\ch_S$ can be represented as 
\begin{equation}
\ch_S \,\simeq \,\oplus_J \ch_J \,\simeq \,\oplus_J \,
\cc_J \otimes \cd_J \,\simeq \,\oplus_J \,\cmplxs^{n_J}\otimes 
\cmplxs^{d_J} \:,
\label{stsp}
\end{equation}
with $n_J, d_J \in \nintg$, $\sum_J n_J d_J=d$, and the action of 
the decoupling group algebra and its commutant given by 
$\widehat{\cmplxs \cg} \simeq $ $\oplus_J \openone_{n_J} \otimes 
\text{Mat}_{d_J} (\cmplxs)$,
$ \widehat{\cmplxs \cg}$$' \simeq $ $\oplus_J \text{Mat}_{n_J} (\cmplxs)
\otimes \openone_{d_J}$, respectively. Because both 
$\Pi_{\widehat{\cg}}(S_\alpha)$ and $\cq_{\widehat{\cg}}(\Delta H_c)$
are in $\widehat{\cmplxs \cg}$$'$, $\cd_J$-subsystems are noiseless and 
their dynamical generation robust regardless of whether 
$\Delta H_c(t)$ belongs to $\widehat{\cmplxs \cg}$ or not. 
This applies in particular if $\cg$ acts irreducibly on $\ch_S$, in 
which case a robust implementation of maximal decoupling 
is achievable by averaging over a 
nice error basis on $\cmplxs^d$~\cite{viola:qc1999a,wocjan:qc2001a}. 
In fact, encoding into $\cd_J$-subsystems may be valuable even in 
situations where the assumption that the controls are in 
$\widehat{\cmplxs \cg}$ cannot be met: as 
$\cq_{\widehat{\cg}}(S_\alpha)$ $\in $ $\widehat{\cmplxs \cg}$$'$, 
$\cd_J$-subsystems remain unaffected by the noise. 
Note that for such subsystems, both the implementation of 
the decoupling scheme and the execution of encoded control 
operations are to be effected through fast modulation of Hamiltonians 
along the control cycle~\cite{viola:qc1999b,viola:qc2000b}. 

Whenever $\cq_{\widehat{\cg}}(\Delta H_c)$ originates from faults in 
$\widehat{\cmplxs \cg}$, additional options are viable. If the 
representation $\mu$ is primary, $Z_{\widehat{\cmplxs \cg}} =
{\cmplxs \openone}$, then {\it any} systematic error is effectively 
eliminated, and no encoding is necessary as long as noise suppression
is ensured, that is, $\Pi_{\widehat{\cg}}(S_\alpha)=0$ for all $\alpha$. 
If $\mu$ is not primary, then elements in the center are diagonal over 
each irreducible component. Thus, encodings into either a 
$\ch_J$-subspace or a $\cc_J$-subsystem are insensitive to the 
control faults and protected against the noise generator if 
$\Pi_{\widehat{\cg}}(S_\alpha) \in Z_{\widehat{\cmplxs\cg}}$ as well. 
In practice, choosing a $\cc_J$-subsystem may be especially appealing,
because not only is universal encoded control achievable by 
less-demanding, slow application of Hamiltonians in 
$\widehat{\cmplxs \cg}$$'$~\cite{viola:qc1999b}, but added robustness 
against {\it arbitrary} control errors in 
$\widehat{\cmplxs \cg}$ is automatically provided~\cite{viola:qc2000b}.
Next, we outline some applications relevant to QIP. 

{\it Example 1: Eulerian Carr-Purcell decoupling on a qubit.-} 
Consider a single decohering qubit, 
$\{ S_\alpha\}=$ $\{ \sigma_z\}$ \cite{viola:qc1998a}. The  
decoupling group $\cg=\cz_2=\{0,1\}$ is represented in 
$\cu(\cmplxs^2)$ as $\widehat{\cg}=\{ \openone, \sigma_x \}$. 
There is one generator, $\gamma_1=1$, hence $L=2$ with no 
overhead with respect to the b.b. case. Let $u_x(s)=$$
\ct \hspace{-.4mm}\exp(\hspace{-.6mm}-i \int_0^{s} 
\hspace{-.5mm}dt' \, h_x (t') )$, for a Hamiltonian 
$h_x (t)\in \widehat{\cmplxs \cg}$ realizing 
$\hat{\gamma}_1=\sigma_x$ (\ref{gen}). If $h_x (t) = f(t) \sigma_x$ 
for some $f(t)$, any choice such that $|\int_0^{\Delta t}$
$ds f(s)|$$=\pi/2$ is acceptable.  On $G({\cg}, {\Gamma})$ choose 
$\cp_E=$ $(\gamma_1,\gamma_1)$. Then 
Eulerian decoupling is accomplished by letting $U_c(t)=u_x(t)$, 
for $t\in [0,\Delta t)$, and $U_c(t)=u_x(s)\sigma_x$ for 
$t\in [\Delta t, \Delta t+s)$, $s\in [0,\Delta t)$. 
By explicit calculation of $\cq_{\widehat{\cg}}(\Delta H_c)$, 
one sees that systematic errors along $\sigma_y,\sigma_z$
produce no effect.  Elimination of residual control errors in 
$Z_{\widehat{\cmplxs\cg}}$ requires using the full Pauli group.

{\it Example 2: Eulerian Pauli decoupling on qubits.-} Let  
$\widehat\cg=\{ \openone, $ $X, $ $ Y, Z \}$ be the Pauli error 
basis for a qubit, with $X=\sigma_x$, $Z=\sigma_z$, and $Y=XZ$. 
This corresponds to $\cg =\cz_2$$\times \cz_2$, 
projectively represented in $\cu(\cmplxs^2)$. $\cg$ has two
generators, {\it e.g.} $\gamma_1=(0,1)$, $\gamma_2=(1,0)$, 
realized as $\hat{\gamma}_1=X$, $\hat{\gamma}_2=Z$, 
respectively.  An Eulerian path on $G(\cg,\Gamma)$ is
$\cp_E=(\gamma_1, \gamma_2, \gamma_1, \gamma_2, 
\gamma_2, \gamma_1, \gamma_2, \gamma_1)$, of length $L=8$. 
The assumption that both $h_\lambda$ and $\Delta h_\lambda$, 
$\lambda=1,2$, are in $\widehat{\cmplxs \cg}$ is automatically 
satisfied, as $\widehat{\cmplxs \cg}=\text{Mat}_2(\cmplxs)$. Then 
(\ref{euler}) results into a robust implementation of maximal averaging, 
$\Pi_{\widehat{\cg}}(\sigma_u)=0$, $u=x,y,z$. For $n$ qubits, 
$\cg=\cz_d$$\times \cz_d$, with $d=2^n$. Thus, $|\cg|$ = $4^n$ 
and since 2 generators are needed for each qubit, 
$L=n 2^{2n+1}$, causing the procedure to be (as in the b.b. 
limit~\cite{viola:qc1999a}) inefficient.

{\it Example 3: Eulerian collective spin-flip decoupling.-}
For $n$ qubits, let $\cg =\cz_2$$\times \cz_2$ 
act via the $n$-fold tensor power representation in 
$\cu((\cmplxs^2)^{\otimes n})$, which is projective 
for $n$ odd, and regular for $n$ even. 
For any $n$, $\widehat\cg=\{ $$\openone, $ $X, $ $ Y, Z \}$, 
where $X=\otimes_{k=1}^n \sigma^{(k)}_x$, $Z=\otimes_{k=1}^n 
\sigma^{(k)}_z$, and $Y=XZ$.
Decoupling according to $\cg$  averages out arbitrary 
linear noise, $\Pi_{\widehat{\cg}}(S_\alpha)=0$, 
$S_\alpha \in$ span$\{ \sigma^{(k)}_u\}$~\cite{viola:qc1999b}. 
For Eulerian implementation, the same path of Example 2 may be used, 
under the appropriate realization of the collective generators 
$\hat{\gamma}_1=X$, $\hat{\gamma}_2=Z$. 
Ensuring $\cg$-symmetrization requires that the control 
Hamiltonians $h_{1,2}(t)\in \widehat{\cmplxs \cg}$. 
Because both $\widehat{\cmplxs \cg}$$'$ and $Z_{\widehat{\cmplxs\cg}}$ 
are non-trivial, residual control errors 
arise due to $\cq_{\widehat{\cg}}(\Delta H_c)$. 
The situation is simpler for $n$ even, as ${\widehat{\cmplxs\cg}}$ 
is abelian hence supporting four $(n-2)$-dimensional irreducible 
subspaces $\ch_J$. Besides being noiseless in the decoupling limit
and insensitive to arbitrary control errors in ${\widehat{\cmplxs\cg}}$,
encoding into a $\ch_J$-subspace is further motivated by the 
possibility to achieve encoded universality via slow application 
of two-body Hamiltonians in 
${\widehat{\cmplxs\cg}}$$'$~\cite{viola:qc2000b}. For $n$ odd, both 
$\cc_J$ and $\cd_J$ factors may occur. Leaving aside details here, we
note that $\cd_J$-subsystems may be useful if implementing 
$\hat{\gamma}_1, \hat{\gamma}_2$ via Hamiltonians in 
${\widehat{\cmplxs\cg}}$ is difficult in practice.

{\it Example 4: Eulerian symmetric decoupling.-} Let $\cg=\cs_n$ 
be the symmetric group of order $n$, acting on 
$\ch_S$ $\simeq$ $(\cmplxs^2)^{\otimes n}$ via 
$\hat{g}_j \otimes_{k=1}^n $ $| \psi_k\rangle =$
$\otimes_{k=1}^n |\psi_{g_j(k)}\rangle$, $g_j \in \cs_n$. In 
particular, the action corresponding to a transposition $(k-1\, k)$,
$k \in \{1,\ldots,n\}$, effects an exchange gate between qubits 
$k-1,k$, denoted by {\tt swap}$_{k-1,k}$. 
Symmetric decoupling allows, in principle, to engineer collective 
error models on $S$ starting from arbitrary linear interactions between
$S$ and $E$ \cite{zanardi:qc1999a,viola:qc2000b}. A minimal
generating set for $\cs_n$ is given by
$\gamma_1=(1\,2)$, $\gamma_2=(1\,2\,\ldots n)$ {\it i.e.}, an 
adjacent transposition and the cyclic shift, respectively. 
$\widehat{\cmplxs \cs_n}$ contains the Heisenberg couplings 
$h(k,l)=\vec{\sigma}_{k}\cdot \vec{\sigma}_{l}$. In fact, every operator
in $\widehat{\cmplxs \cs_n}$ can be realized by applying 
Heisenberg Hamiltonians \cite{kempe:qc2001a}. Focus, for instance, 
on $\cs_3$-symmetrization, which may be relevant for inducing 
collective decoherence on blocks of 3 qubits \cite{wu:qc2002a}. 
Then $\hat{\gamma}_1=$ {\tt swap}$_{1,2}$ and 
$\hat{\gamma}_2=${\tt swap}$_{1,2}$ {\tt swap}$_{2,3}$, with 
$L=12$. Because $\exp(-i \pi h(k,l)/4)$ 
={\tt swap}$_{k,l}$, $\hat{\gamma}_1$ can be implemented
by choosing $h_1=a_1 h(1,2)$, with strength 
$a_1=\pi/4\Delta t$, while $\hat{\gamma}_2$ can be realized by a 
piecewise-constant Hamiltonian $h_2(t)=a_{2}$$ h(2,3)$ for $t\in 
[0,\Delta t/2)$, $h_2(t)=a_{2} $$h(1,2)$ for $t\in [\Delta t/2, 
\Delta t]$, $a_{2}=\pi/2\Delta t$. A Eulerian path on 
$G({\cs_3},{\Gamma})$ is $\cp_E$ 
$=({\gamma}_2, {\gamma}_2, $ ${\gamma}_2, {\gamma}_1,$
${\gamma}_2, {\gamma}_1, $ ${\gamma}_1, {\gamma}_2,$
${\gamma}_1, {\gamma}_1, $ ${\gamma}_2, {\gamma}_1)$.
Eulerian decoupling (\ref{euler}) then allows for a robust dynamical 
generation of the smallest non-trivial noiseless 
subsystem \cite{knill:qc2000a,viola:qc2001a}, 
supported by a factor $\cd_J\simeq 
\cmplxs^2$ carrying the two-dimensional irreducible component 
$J=[2\,1]$ of $\cs_3$.  

{\it Conclusion.-} We developed an approach to dynamical decoupling
that combines the group-theoretical essence of the b.b.  setting 
with graph-theoretical control design according to Eulerian
cycles.  Besides allowing for considerable leeway in the physical
implementation of the basic control generators, Eulerian decoupling
eliminates the need for unfeasible b.b. pulses and naturally 
incorporates robustness against realistic control faults. Combined 
with quantum coding techniques, our results significantly improve the
prospects that dynamical decoupling becomes a practical tool for
reliably controlling quantum systems and quantum information.

\acknowledgments 
Supported by the DOE (contract W-7405-ENG-36) and by the NSA. 
L.V. also gratefully acknowledges support from a J.R. 
Oppenheimer Fellowship.


\begin{thebibliography}{22}
\bibitem[{not({\natexlab{a}})}]{note}
\bibinfo{note}{For a projective representation, $\widehat{g_j
  g_k}=\omega(g_j,g_k) \hat{g}_j \hat{g}_k$ for some phase $\omega(g_j,g_k)$,
  hence $\widehat{\cg}$ is not a group in general.}
\bibitem[{not({\natexlab{b}})}]{note2}
\bibinfo{note}{Here, the set of operators $\widehat{\cg/\cg_0}$ is defined with
  respect to an arbitrary choice of coset representatives of $\cg_0$ and is not
  necessarily a projective representation.}
\end{thebibliography}
\end{document}